\documentclass[doublecol]{epl2} 

\usepackage{xcolor}
\usepackage{ulem}

\def\gsim{\mathrel{\rlap{\lower4pt\hbox{\hskip1pt$\sim$}}
    \raise1pt\hbox{$>$}}}       

\def\be{\begin{equation}}
\def\bea{\begin{eqnarray}}
\def\ee{\end{equation}}
\def\eea{\end{eqnarray}}

\def\z{\vert 0 \rangle}

\def\o{\vert 1 \rangle}

\def\Tr{{\rm Tr}}

\def\gsim{\mathrel{\rlap{\lower4pt\hbox{\hskip1pt$\sim$}}

    \raise1pt\hbox{$>$}}}       

\def\gsim{\mathrel{\rlap{\lower4pt\hbox{\hskip1pt$\sim$}}
    \raise1pt\hbox{$>$}}}       

\def\z{|0\rangle}
\def\o{|1\rangle}

\title{Black Hole Information, Replica Wormholes, and Macroscopic Entanglement} 

\author{Xavier Calmet \thanks{E-mail: \email{x.calmet@sussex.ac.uk}}\inst{1} \and Stephen~D.~H.~Hsu\thanks{E-mail: \email{hsusteve@gmail.com}} \inst{2}}
\shortauthor{X. Calmet and S.D.H. Hsu}

\institute{                    
  \inst{1} Department of Physics and Astronomy,\\
University of Sussex, Brighton, BN1 9QH, United Kingdom\\
  \inst{2} Department of Physics and Astronomy\\ Michigan State University, East Lansing, Michigan 48823, USA
}


\abstract{
In this invited review, we discuss the evaporation of a black hole, with emphasis on the resulting macroscopically distinct patterns of Hawking radiation. The density matrix of this radiation can approach a pure final state in the form of a highly entangled macroscopic superposition state. We note that this exact property is exhibited in replica wormhole calculations, and also in quantum hair effects on Hawking radiation amplitudes. Finally, we revisit the information paradox (Mathur's theorem and firewalls), showing that it can be resolved by macroscopic entanglement.}

\begin{document}

\maketitle

\section{Introduction: Black Hole Information}

Nearly 50 years ago, Hawking argued that that black holes cause pure states to evolve into mixed states \cite{Hawking:1976ra,Hawking1976}. Based on the causal properties of (classical) black hole spacetimes, he concluded that quantum information that falls into a black hole does not re-emerge in the form of radiation. Thus, the final radiation state left after a black hole has fully evaporated must be a mixed state, even if the hole was originally formed in a pure state.

This article is a invited review of some important, but poorly understood, aspects of black hole information. It is based on work in \cite{Calmet:2021cip,Calmet:2021stu,Hsu:2013cw,Hsu:2013fra,Calmet:2024tgm,Calmet:2023gbw,Calmet:2019eof,Calmet:2022swf}, and lectures presented at a number of universities. The article is pedagogical and is meant to be accessible to students and non-specialists. Experts should consult the original literature cited above for technical details.

We emphasize that a pure final radiation state, resulting from unitary evolution of the black hole initial state, must be a macroscopic superposition state. Previous analyses of the information problem have overlooked the possibility of entanglement between these macroscopically different states. We present evidence for this entanglement, and that it allows a pure final radiation state.

\section{Phenomenology of Black Hole evaporation: recoil trajectories}

In our discussion we consider black holes much larger than the Planck mass, so that the spacetime geometry is classical to good approximation, and quantum effects are small corrections. Some important properties of the evaporation:

\smallskip
1. Emitted Hawking quanta have low energy and long wavelength, of order the size of the black hole

\smallskip
2. Evaporation takes place very slowly, with (on average) one quantum emitted per light crossing time of the hole

\smallskip
3. The spacetime curvature near the horizon of the black hole is small, and a freely falling observer should observe the flat space vacuum state

\smallskip
These properties hold because the black hole temperature (hence, typical wavelength of emitted quanta), rate of radiation emission, and the deviation of spacetime at the horizon from flat space all vanish with some power of the black hole mass in Planck units.

Energy-momentum conservation requires that the black hole recoil and shrink slightly with each emitted quantum. The hole appears to execute an approximate random walk while it evaporates. The long timescales between emissions allow a build up of displacement, so that by the time a significant fraction of its mass has evaporated, the uncertainty in its position is macroscopic: $\Delta x \sim M^2$, which is enormously larger than its original Schwarzschild radius, $r \sim M$.

Each radiation emission is described by a quantum amplitude, and {\it because quantum mechanics is linear}, the quantum state describing multiple emissions is a superposition of many possible radiation states. Consequently, the state describing the hole itself is a superposition of many macroscopically different recoil trajectories. 

Visualization of the black hole recoil trajectory provides an intuitive picture of why the usual formulations of the information paradox are incomplete: they assume a single spacetime background for the entire evaporation. Hence, entanglement {\it across} different components of the superposition state (corresponding to different recoil trajectories) are never considered.

\section{The Hilbert space of Hawking radiation}

Consider the entirety of the radiation emitted by the black hole during its evaporation. For example, if all emitted particles are massless, this radiation occupies an expanding shell with thickness equal to the black hole lifetime. Long after the black hole is gone, this radiation pattern continues to propagate toward the asymptotic boundary of spacetime. The density of individual quanta in this expanding shell is very low, except for the last burst of radiation emitted when the hole is very small and hot. 

We can consider the Hilbert space of all possible states of this radiation pattern. It is subject to a constraint on total energy, momentum, and conserved charges. But any pattern that satisfies the constraints is a possible state in the Hilbert space, and in general they correspond to different recoil trajectories of the black hole.

Given a specific spatial coarse graining scale, we can define macroscopically different patterns of states. Modulo this coarse graining, it is clear that the number of macroscopically different states in our Hilbert space is almost as large as the total radiation Hilbert space. In particular, its dimension grows like the exponential of the original area of the black hole.  

While the uncertainty $\Delta x$ in the position of the hole only grows like $\sim M^2$, the number of macroscopically different recoil trajectories clearly grows exponentially in the area, since they are in one to one correspondence with the set of macroscopically distinct radiation patterns.

\section{Density Matrix}

Now consider the density matrix $\rho$ describing the Hawking radiation state at late times. The relevant Hilbert space is the one described in the previous section.

It has been known for some time that a large number of exponentially small corrections to even a maximally mixed $\rho$ can produce an order one correction to the purity of the density matrix  \cite{Hsu:2013cw,Hsu:2013fra,Papadodimas:2012aq}. At leading order, the maximally mixed density matrix is $\rho_{ij} \sim d^{-1} \, \delta_{ij}$, where $d$ is the dimensionality of the Hilbert space. Now include corrections to the density matrix: 
\begin{equation}
\rho_{ij} \sim d^{-1} \, \delta_{ij} + \delta \rho_{ij}~~. 
\end{equation}
The leading order contribution to $\Tr [ \rho^2 ]$ is 
$$d^{-2} \, \Tr \, \delta_{ij} = d^{-1}~.$$ 
Consider $\Tr [ \delta \rho^2 ]$, where $\delta \rho$ is $d^{-1}$ times a non-sparse matrix whose entries are of order $x$. By non-sparse we mean that most or all of the entries of the matrix $\delta \rho$ are non-zero (including off-diagonal entries), reflecting widespread entanglement between radiation states. (This widespread entanglement is indicated by the wormhole configurations we discuss later, which connect any $i,j$ pair.) Here $x$ is the characteristic size of subleading corrections resulting from quantum gravitational effects. Due to the non-sparse matrix structure, multiplication of $\delta \rho$ with itself can produce entries in $\delta \rho^2$ of order $d^{-1} x^2$, so that $\Tr [ \delta \rho^2 ]$ can be as large as $x^2$. 

In earlier work on black hole evaporation the origin of the small corrections $x$ was not known. Non-perturbative effects in quantum gravity are expected to have size of order $\exp (- S)$, where $S$ is of order the black hole entropy, but no method of calculating these effects was known.

\begin{figure}
\centering
\includegraphics[scale=.55]{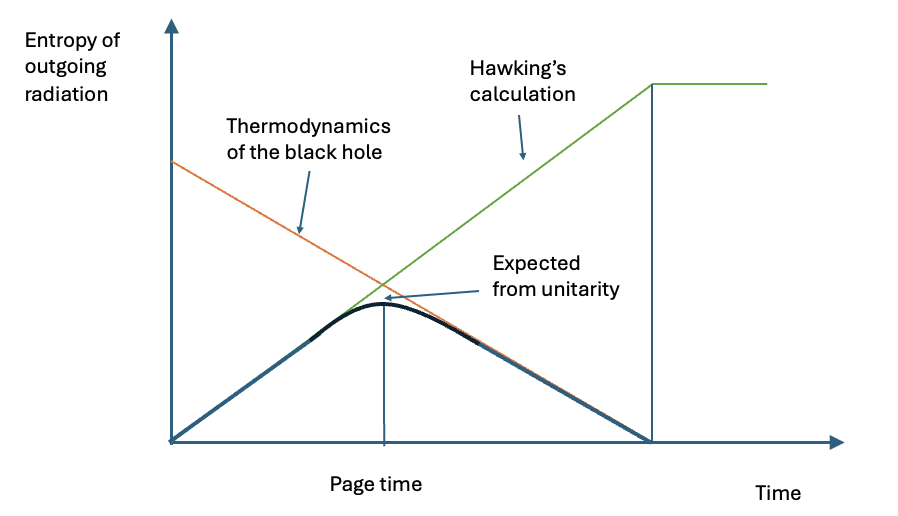}
\caption{The entanglement entropy of the radiation state as a function of time. In the leading order Hawking calculation the entropy increases monotonically. The Page curve (black) assumes unitary evolution of the black hole and requires order one corrections to the Hawking calculation while the hole is still macroscopic - i.e., at the Page time. Wormhole calculations provide explicit examples for how this occurs.}
\label{pagecurve}
\end{figure}

Now consider the evaporation of the black hole in the presence of quantum gravity corrections $x$, such as those produced by replica wormholes. The dimensionality $d$ of the radiation Hilbert space increases as more quanta are emitted, and the entropy (area) of the black hole decreases, so that the size of quantum gravity corrections, $x \sim \exp( -S )$ increases with time. At an intermediate time (i.e., the Page time), the growing $x^2$ correction to the purity dominates the Gibbons-Hawking contribution $1/d$, which is shrinking. After this time the purity is increasing toward unity, rather than decreasing with time, in accordance with the Page curve \cite{Page:2013dx}.

\section{Replica wormholes}

In \cite{Calmet:2024tgm} we discuss recent applications of the Euclidean path integral formulation of quantum gravity to the black hole information problem \cite{Penington:2019npb,Almheiri:2019psf,Almheiri:2019qdq,Almheiri:2019yqk,Penington:2019kki,Gautason:2020tmk,Hartman:2020swn}, see, e.g., \cite{Almheiri:2020cfm} for a recent review. Our main interest is in what these results suggest about the real time physical mechanism behind unitary evaporation of black holes. In other words, what is the mechanism by which the Hawking radiation state becomes a pure state? 

There are many open questions regarding the Euclidean path integral 
formulation of quantum gravity, which remain unresolved \cite{Unruh:1988is,Kiefer}. However, the recent results reviewed below follow mainly from the assumption that certain wormhole configurations connecting black hole interiors are the next to leading correction to the Gibbons-Hawking saddlepoint describing a (Euclideanized) black hole.

Let $\Psi$ denote the initial state of the black hole, and $i,j$ label specific radiation states that result from Hawking evaporation. Then $\rho_{ij} = \langle i \vert \Psi \rangle \langle \Psi \vert j \rangle$ is the density matrix in the $i,j$ basis. Define the purity of the density matrix by $\Tr[\, \rho_{ij}^2 \, ]$, with pure states satisfying $\Tr[\, \rho_{ij}^2 \, ] = 1$.

Now consider (Figure 2) two contributions to a Euclidean path integral expression for $\Tr[\, \rho_{ij}^2 \, ]$, from the Gibbons-Hawking configurations and from wormhole configurations which connect the interiors of the black holes \cite{Almheiri:2020cfm}. The former have smaller Euclidean action, but the number of wormhole configurations grows as the square of the dimensionality $d$ of the radiation Hilbert space. This is because a wormhole configuration can connect the interiors of any two different black holes $i,j$.

At early stages of evaporation, when $\log d$ is small compared to the area $A$ of the evaporating hole, the Gibbons-Hawking saddlepoint dominates: $\Tr[\, \rho_{ij}^2 \, ]  \ll 1$, and purity decreases with each emission. However as the black hole continues to evaporate, the wormhole contribution begins to dominate. Late in the evaporation process ($\log d > A$), $\Tr[\, \rho_{ij}^2 \, ]$ increases, potentially approaching unity, indicating a pure final state. The transition between dominance of the Gibbons-Hawking configuration and the wormhole configurations is expected to occur at the Page time \cite{Almheiri:2020cfm,Almheiri:2019hni}.

In some low-dimensional models the replica wormhole calculation can be carried out explicitly, with the desired results. Further, these calculations reproduce the island formula involving quantum extremal surfaces (related to the Ryu-Takayanagi formula \cite{Ryu:2006bv}) \cite{Penington:2019npb,Almheiri:2019psf,Almheiri:2019yqk}. However, this does not directly reveal the physical mechanism (i.e., in real time) that allows the Hawking radiation states to encode the black hole quantum state $\Psi$.

\begin{figure}
    \centering
    \includegraphics[width=1\linewidth]{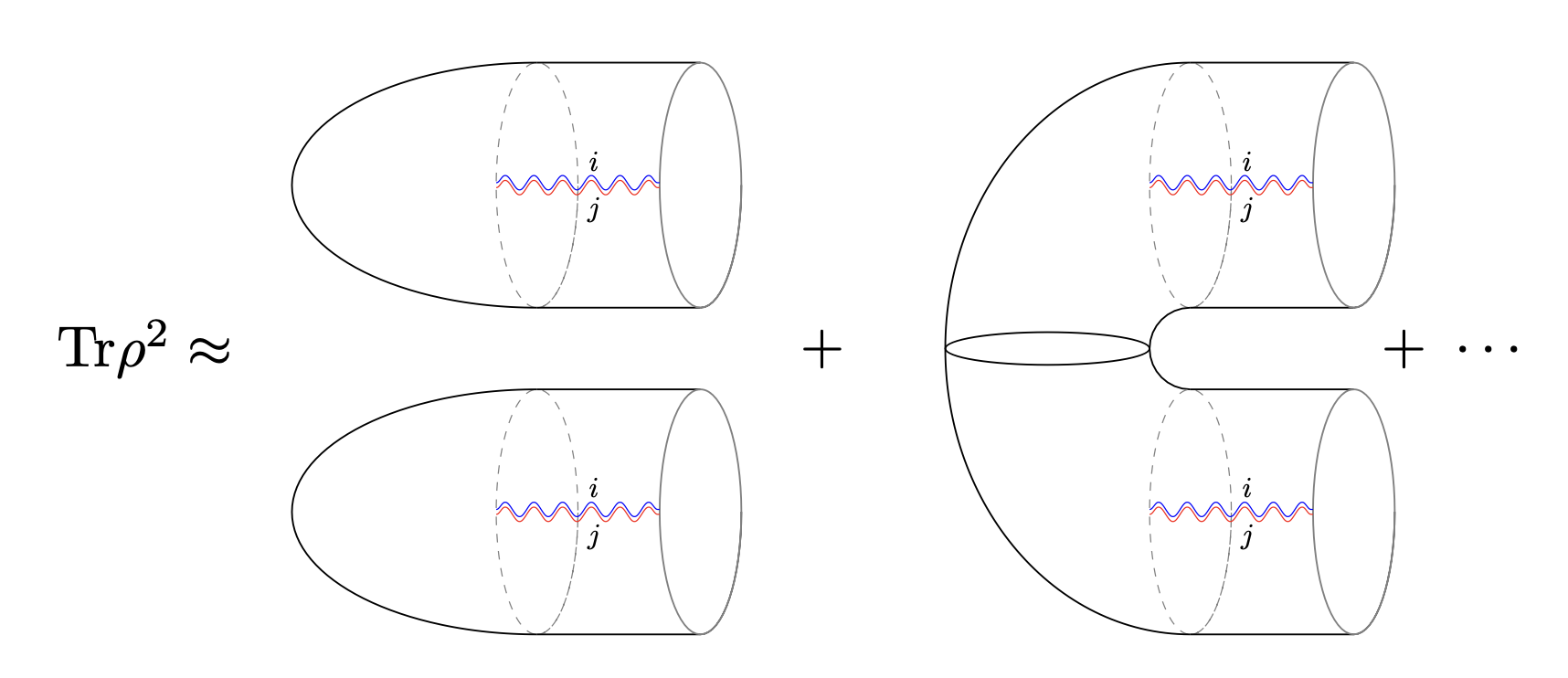}
    \caption{Two saddlepoints which contribute to the Euclidean path integral expression for the purity $\Tr\, \rho^2 \,$ in quantum gravity. On the left, the Gibbons-Hawking saddlepoint constructed from Euclideanized black hole spacetimes. On the right, the wormhole configuration connecting black hole interiors.}
    \label{saddlepoints}
\end{figure}
 
The wormhole calculation only involves long wavelength, low energy physics. It does not rely on a specific UV completion of quantum gravity. Hence the corresponding real time physical picture is likely to depend only on the long wavelength properties of quantum gravity. These should be manifest in effective field theory, which has been used to demonstrate the existence of quantum hair \cite{Calmet:2021stu,Calmet:2021cip,Calmet:2019eof,Calmet:2022swf,Calmet:2023gbw}. Quantum hair allows Hawking evaporation amplitudes to depend on the internal state of the black hole, as we discuss in section 3.

The combinatorial property that allows the wormhole configurations to dominate the path integral at large $d$ (late in the evaporation process) reflects the fact that a wormhole can connect {\it any} two black hole interiors. Specific wormhole solutions only depend on the semiclassical properties of the interiors.
\begin{figure}
\centering
\includegraphics[scale=0.5]{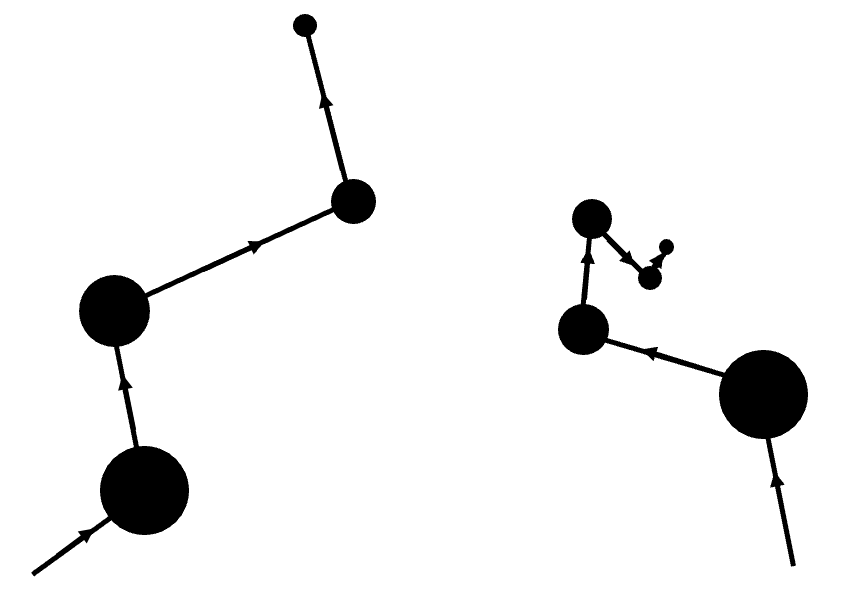}
\caption{Two possible recoil trajectories of the black hole as it evaporates. The number of such trajectories is, modulo a choice of coarse graining, in 1-1 correspondence with the set of radiation patterns. Therefore the number of distinct trajectories is exponential in the initial area of the black hole. Each recoil trajectory corresponds to a distinct background spacetime, and the set of coordinate transformations required to place the black hole at the origin of the Penrose diagram is different for each trajectory.}
\label{recoil1}
\end{figure}

\begin{figure}
\centering
\includegraphics[scale=0.6]{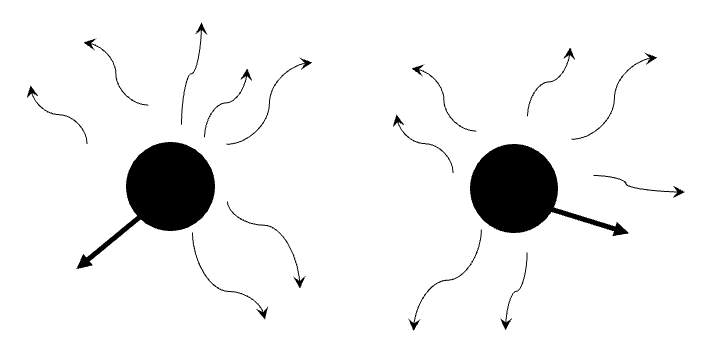}
\caption{Each radiation pattern, due to conservation of energy-momentum-spin, corresponds to a different black hole recoil trajectory. The radiation state at late times is a superposition of macroscopically different patterns. Consequently, the quantum gravity state describing the evaporation is a superposition of macroscopically different spacetimes - one for each distinct recoil trajectory.}
\label{recoil2}
\end{figure}
As we discussed earlier, different radiation states $i,j$ generally correspond to macroscopically different recoil trajectories of the black hole \cite{Page:1979tc}. At late times the uncertainty in the center of mass position of the hole is of order its initial mass in Planck units: $\Delta x \sim M^2$, which is enormously larger than the original Schwarzschild radius $R \sim M$. Different recoil trajectories, i.e., distinct spacetime backgrounds, are typically subsumed into a single Penrose diagram for purposes of visualization, but in fact each radiation pattern leads to a different center of mass position, recoil momentum, and precise internal state of the hole as time progresses. The coordinate transformation required to place the black hole at the origin of the Penrose coordinates is different for each recoil trajectory.

However, the semiclassical geometry of the black hole interior should be roughly independent of the center of mass position of the hole, its velocity, and microscopic aspects of the state of the emitted radiation. For example, two typical black holes located at $x_1$ and $x_2$ due to different recoil trajectories nevertheless have similar semiclassical internal geometry, and can be connected by a wormhole whose action is independent of $( x_1 - x_2 )$. We expect that the wormhole dynamics (i.e., Euclidean action of a specific solution) is, to leading order, invariant under the specific nature of typical $i,j$ states. This property is explicit in low-dimensional calculations \cite{Penington:2019npb,Almheiri:2019psf,Almheiri:2019yqk}.

Thus, the replica wormhole picture, in which large numbers of $i,j$ pairs overcome exponential suppression from larger Euclidean action, {\it implies} that $\rho_{ij}$ at late times describes a superposition state of macroscopically different spacetimes. Each black hole recoil trajectory corresponds to a different spacetime, but wormholes connect any pair of them via the black hole interiors. This has important consequences for firewall and monogamy of entanglement constructions, as we discuss below.

In some proposed interpretations of the wormhole results, it is suggested that modes in the entanglement wedge (i.e., inside the black hole) be {\it identified} with radiation modes. This requires not just nonlocality across the horizon (i.e., perhaps the Euclidean wormholes from the path integral calculation indicate the presence of wormholes in Minkowski spacetime that transport information across the horizon), but something even stronger: a kind of holographic identification of modes inside and outside the hole \cite{Maldacena:2013xja}. Below we note that quantum hair allows radiation modes produced in Hawking evaporation to {\it encode} the interior information while still remaining independent degrees of freedom. 

Both replica wormhole and quantum hair approaches imply that radiation states $\rho_{ij}$ are entangled macroscopic superpositions of different spacetimes. As we explain later, this {\it invalidates} firewall and monogamy of entanglement constructions which play an important role in consideration of physical mechanisms for purification of the radiation state. 

While replica wormholes provide a plausible means to compute the quantum gravity effects $x$ which purify the density matrix $\rho$, we still lack a real time physical understanding of this process. We will next discuss what low energy effective field theory applied to Lorentzian quantum gravity tells us about black hole evaporation.

\section{Macroscopic Entanglement from Quantum Hair}

Calculations using the Vilkovisky-DeWitt effective field theory of quantum gravity show that the quantum state of the external gravity field depends on the internal state of the black hole \cite{Calmet:2021stu,Calmet:2021cip,Calmet:2019eof,Calmet:2023gbw,Calmet:2022swf}. The causal structure of the {\it classical} black hole spacetime does not govern these quantum effects, which we refer to as quantum hair. Quantum hair affects Hawking evaporation, allowing radiation amplitudes to depend on the internal state of the hole \cite{Calmet:2021stu,Calmet:2021cip,Calmet:2019eof,Calmet:2022swf,Calmet:2023gbw}. 

The example that can be treated most explicitly is that of a spherically symmetric, pressureless dust ball that collapses to form a black hole. In that specific example, we can use the  Vilkovisky-DeWitt effective field theory of quantum gravity to compute corrections to the exterior metric. These corrections depend on the density profile of the dust ball. In classical general relativity, Birkhoff's theorem and spherical symmetry imply that the classical metric can only depend on the total mass and cannot depend on the density profile. This is the simplest example of information about the interior of the hole that is available outside the horizon, due entirely to quantum gravity effects. It has further been demonstrated that these quantum corrections influence Hawking radiation amplitudes \cite{Calmet:2023gbw}. 

One can also make more general arguments which suggest that the entire quantum state of a compact object is manifested in the quantum state of the gravity field sourced by that object \cite{Raju:2020smc}.

Under the assumption that Hawking radiation amplitudes, via quantum hair, depend on the internal black hole state, we can use simple quantum mechanics to construct the final pure radiation state as a function of the original black hole state $\Psi$. The radiation state is a linear function of the black hole state, and $\Psi$ can, in principle, be reconstructed from the radiation state.

However, the radiation state is a complex macroscopic superposition state. The development of this state over time cannot be described with reference to a single background geometry. As we mentioned in the last section, this is also a feature of the path integral computation using replica wormholes.   

Let $\psi_g (E)$ be the asymptotic state of the gravity field sourced by a compact object (the black hole), which is an energy eigenstate with eigenvalue $E$. Each distinct energy eigenstate of the compact source produces a different quantum state of its gravity field. There are $\sim \exp S$ such eigenstates. (We make the plausible assumption that there are no accidental degeneracies in the energy spectrum. In the presence of such, the Hawking amplitudes given below can depend on more complicated aspects of the quantum state.)

A semiclassical matter configuration is a superposition of energy eigenstates 
\begin{equation}
\Psi = \sum_n c_n \psi_n~~,   
\label{Psi}
\end{equation}    
with support concentrated in some narrow band of energies (i.e., reflected in the coefficients $c_n$). Here $\psi_n$ are energy eigenstates with eigenvalues $E_n$.
The resulting gravity state $\psi_g$ (i.e., the state of the gravity field sourced by the semiclassical object above) is itself a superposition state: 
\begin{equation}
\psi_g = \sum_n c_n \psi_g (E_n) \equiv \sum_n c_n ~\vert \, g(E_n) \, \rangle
\end{equation} 

When the black hole emits the first radiation quantum $r_1$ the state describing its {\it exterior} evolves into the state on the right, below:
\begin{equation}
    \sum_n c_n ~\vert \, g(E_n) \, \rangle ~\rightarrow~ \sum_n \sum_{r_1} c_n ~ \alpha (E_n, r_1) ~ \vert \, g(E_n - \Delta_1), r_1 \rangle 
    \label{rad1} ~~. 
\end{equation}
Here $\alpha (E, r)$ is the amplitude for a black hole in energy eigenstate $E$ to emit a radiation quantum with quantum numbers $r$. We suppress other quantum numbers of the hole, such as its charge, angular momentum, etc. $\alpha (E, r)$ is computable, at least in principle. It depends both on the state $r$ of the emitted quantum and of the black hole through its energy eigenvalue $E$. The amplitude can depend on $E$ because of quantum hair. Explicit computations demonstrate that, in violation of classical no hair theorems, features of the black hole internal state are manifested in the quantum hair \cite{Calmet:2021stu,Calmet:2021cip,Calmet:2019eof,Calmet:2022swf}.

In this notation $g$ refers to the exterior geometry and $r_1$ to the radiation. The next emission leads to
\begin{eqnarray}
\label{rad2}
    \sum_n \sum_{r_1,r_2} c_n ~ \alpha (E_n, r_1) ~ \alpha (E_n - \Delta_1, r_2) ~ \nonumber \\
    \times ~~ \vert \, g(E_n - \Delta_1 - \Delta_2), r_1, r_2 \rangle 
\end{eqnarray}
and the final radiation state is
\begin{eqnarray}
\label{rad3}
\sum_n \sum_{r_1,r_2,\ldots,r_N} c_n ~\alpha(E_n, r_1) ~\alpha(E_n - \Delta_1, r_2) ~ 
\nonumber \\
~ \times ~~
\alpha(E_n - \Delta_1 - \Delta_2, r_3 ) ~
\cdots ~ \vert \, r_1 ~ r_2 ~ \cdots ~ r_N \rangle ~. 
\end{eqnarray} 

Compare this to the radiation produced by thermal evaporation. At each step, the emission amplitudes differ from thermal by a small amount, because they can depend on the specific internal state of the hole (and not just on the quantities corresponding to classical hair). These corrections are small - they are suppressed by $S^{-k}$ and $\exp(-S)$, where $S$ is the entropy of the hole. These factors arise, respectively, from the perturbative suppression of quantum hair in the  Vilkovisky-DeWitt effective field theory calculation, and because energy eigenvalue splittings of nearby states of a black hole are of order $\exp(-S)$. 

The {\it statistical features} of the radiation pattern (distribution of energy densities, etc.) and the black hole recoil trajectory (e.g., uncertainty $\Delta x (t)$ in center of mass position as evaporation proceeds) corresponding to (\ref{rad3}) differ only slightly from the case of random thermal Hawking emission. The difference is that, due to quantum hair and the resulting coherent amplitudes $\alpha (E, r)$, the state (\ref{rad3}) is related to the original state of the black hole $\Psi$ via a unitary transformation - indeed, the final state is fully determined by the coefficients $c_n$ from (\ref{Psi}).

At intermediate stages of the evaporation, the internal state of the black hole appears in equations 
(\ref{rad1})--(\ref{rad2}), via the external gravity state $\vert \, g(E_n) \, \rangle$, which depends on the black hole energy $E_n$. Tracing over this information yields a mixed density matrix $\rho$ which has the non-sparse property discussed in Section 2: it connects all possible radiation states $i,j$, which correspond here to the sum over $r_1, r_2, \ldots$ When the hole has completely evaporated we are left with the pure state in (\ref{rad3}).

The reader may be familiar with results such as Mathur's theorem or firewall constructions that claim to show that small corrections cannot solve the information paradox. Below we will explain why the evolution described above, which is explicitly unitary, is not excluded by these results. The reason is that the radiation state is only pure if one considers {\it all branches} of (\ref{rad3}) - these describe emission of the $k$-th quanta taking place on macroscopically different background spacetimes - i.e., evaporation from black holes with different recoil trajectories - the holes may be in different locations with different velocities when the $k$-th emission occurs! 

The constructions we review below are performed on a single background spacetime and do not contemplate entanglement between different geometries. However, both the quantum hair and replica wormhole results suggest that such entanglement is central to reproducing the Page curve and to restoration of purity.

\section{Mathur theorem and Firewalls}

Mathur's theorem \cite{Mathur:2009hf,Mathur:2011uj} provides the clearest formulation of the information paradox. Subject to its assumptions, the entanglement entropy of the emitted radiation is shown to increase monotonically, even in the presence of small quantum corrections. The main assumption is that the black hole interior is causally disconnected from the region just outside the horizon where the Hawking modes originate. In this picture, after $N$ quanta are emitted we have a state of the form
\begin{eqnarray}
|\Psi\rangle ~\approx~ |\psi\rangle_M &\otimes&     \Big(  \z_{e_1}\z_{b_1}+ \o_{e_1}\o_{b_1}\Big)\cr
&\otimes&  \Big(  \z_{e_2}\z_{b_2}+ \o_{e_2}\o_{b_2}\Big)\cr
&\dots&\cr
&\otimes& \Big( \z_{e_N}\z_{b_N}+ \o_{e_N}\o_{b_N}\Big)~,
\label{six} \nonumber
\end{eqnarray}
where $b_i$ are Hawking modes and $e_i$ are modes that fall behind the horizon. The entanglement between $e,b$ modes leads to increasing entanglement between the interior and modes which escape to infinity: hence, increasing entanglement entropy after tracing over the interior. This leads to the Hawking information paradox once the black hole evaporates completely. In his formulation the initial matter state appears in  a tensor product with all the other quanta, since the pair creation happens far away. 

However, the analysis above is confined to one background spacetime: the $e,b$ modes are {\it defined} as excitations on this background. The possibility of entanglement between these degrees of freedom and other radiation modes which are emitted on {\it other} branches of the state (\ref{rad3}) are not considered.

This differs from the physical picture implied by quantum hair (cf. equations (\ref{rad1}--{\ref{rad3}})), and also by the replica wormhole calculation. As we have emphasized, both suggest that purification of the radiation state results from entanglement between radiation states $i,j$, where $i$ and $j$ may represent different spacetime backgrounds.  

Now let us consider the AMPS firewall construction \cite{Almheiri:2012rt,Almheiri:2013hfa}. The AMPS set-up also considers only a subset of the space of radiation states spanned by $i$ and $j$. On a {\it fixed} spacetime background, they define the following: A = a set of modes in the black hole interior, B = near horizon exterior modes (late time radiation), C = early radiation. Let $X \asymp Y$ denote {\it X strongly entangled with Y}. 

1. The equivalence principle (i.e., normal vacuum fluctuations near the horizon of a large hole) implies $A \asymp B$. This is analogous to Mathur's entangled $e,b$ states above, with $A$ modes originating as $e$ modes, which are highly entangled with outgoing $b$ modes $\in B$. 

2. Purity of the final radiation state requires: $B \asymp  C$. That is, for the late radiation $B$ to purify the early radiation $C$, the two must be highly entangled.

Because $B$ cannot be strongly entangled with {\it both} $A$ and $C$ (due to monogamy of entanglement), we have to give up either the equivalence principle (i.e., normal QFT state near the horizon; deviation from this is the so-called ``firewall'' ), or purity of the final radiation state. 

However, the set of radiation states in our earlier discussion, labeled by $i,j$, is much larger than the set of modes considered by AMPS, which are defined with respect to a fixed spacetime geometry. There is an alternative to assumption 2 above: the radiation state on any {\it fixed} background evolves into a {\it mixed} state. Purity only results when entanglement between all modes, emitted from black holes with different recoil trajectories, are taken into account. This is apparent from equations (\ref{rad1}--\ref{rad3}), since the pure final state involves all such modes, and also suggested by the fact that replica wormholes can connect the interiors of two black holes that have had very different recoil trajectories.

Finally, let us address an erroneous argument that claims the macroscopic superposition aspect of the final radiation state does not affect the firewall construction. The erroneous claim is as follows: 

\begin{quote}
The uncertainty in the center of mass location of the black hole only grows like a polynomial in the mass (i.e., as $M^2$). This implies that only a negligible amount of quantum information is connected to the recoil trajectory of the hole, and it is irrelevant to the firewall construction and the information paradox.
\end{quote}

This claim focuses on the final uncertainty in position of the black hole, but ignores the exponentially many {\it recoil trajectories}. The fact that there are $\sim \exp A$ recoil trajectories is obvious given that the recoils are in direct correspondence to distinct patterns of radiation emission. Two different $k$-th Hawking quanta (emitted with different energy, momentum, mass, charge) cause distinct $k$-th perturbations to the recoil trajectory. While the final uncertainty in the black hole position only grows like a polynomial in its initial mass, the number of paths it could have followed (recoil trajectories) is exponential in its original area.

We discuss recoil trajectories mainly because they provide vivid imagery of the evaporation process. In fact, we could simply observe that macroscopically different radiation patterns must decohere from each other, leading to the possibility of entanglements across the very high dimensional radiation Hilbert space. These have not been considered in earlier formulations of the information paradox.

\section{Complementarity}

In some proposed interpretations of the wormhole results, it is suggested that modes in the entanglement wedge (i.e., inside the black hole) be {\it identified} with radiation modes. This requires not just nonlocality across the horizon (i.e., perhaps the Euclidean wormholes from the path integral calculation indicate the presence of wormholes in Minkowksi spacetime that transport information across the horizon), but something even stronger: a kind of holographic identification of modes inside and outside the hole \cite{Maldacena:2013xja}. In the past, related ideas have also been referred to as black hole complementarity \cite{tHooft:1990fkf,Susskind:1993if}.

Evaporation in the presence of quantum hair leads to equations (\ref{rad1}--\ref{rad3}), in which the radiation state {\it encodes} the original black hole pure state $\Psi = \sum_n c_n \psi_n$. In fact, the final expression for the radiation state (\ref{rad3}) is {\it linear} in the coefficients $c_n$. Measurements of the radiation state can recover the $c_n$ and hence the original state $\Psi$. We see that it is {\it not necessary to identify} the radiation modes with the interior modes in the entanglement wedge. The radiation modes are independent degrees of freedom, but their final state is determined by the initial state $\Psi$ via unitary evolution.

It has been suggested that some kind of complementarity between the experiences of observers who do or do not fall into the hole must play a role in the resolution of the information paradox \cite{tHooft:1990fkf,Susskind:1993if}. The subjective experience of an observer who falls into a black hole must be reconciled with the requirement that an observer at infinity sees a pure final radiation state. We can now see that this complementarity is between observers of two types. One type experiences decoherence (and hence observes the hole on only {\it one} of its exponentially many different recoil trajectories), falls behind the horizon (no drama or firewall), and believes that black hole evaporation leads to a mixed radiation state. Another observer at infinity, who is hypothetically not subject to decoherence, can make precise measurements across the entire radiation Hilbert space (i.e., can detect entanglement across macroscopically different patterns of radiation), and can reach the complementary conclusion that the radiation is in a pure state.

\section{Conclusions}

After 50 years there is a growing consensus that Hawking's original argument that black holes cause pure states to evolve into mixed states is not correct. His calculations did not account for quantum gravitational effects, but those were long thought to be too small to purify the final radiation state. In constructing the paradox, we can restrict analysis to scenarios involving a large, semiclassical black hole. In this context the quantum corrections are expected to be suppressed by powers of the black hole mass, or even exponentially small in its area (i.e., for non-perturbative effects). However, as we have emphasized, the large dimensionality of the radiation Hilbert space (which is exponentially large in the area of the black hole) can enhance the effect of otherwise small corrections to Hawking's approximations. Small entanglements, spread over the entire Hilbert space, can purify the final state.

Recent work using the replica wormhole approximation to the Euclidean path integral suggests that Hawking radiation from a black hole approaches a pure state, preserving unitarity. 

We have emphasized that these results depend on the ability of wormholes to connect black hole interiors corresponding to any pair of typical radiation states $i,j$. This implies that the density matrix $\rho_{ij}$ approaches a superposition of spacetimes that result from radically different black hole recoil trajectories. In other words, purification of the radiation state occurs across many branches of a macroscopic superposition state. We showed that this invalidates both the Mathur \cite{Mathur:2009hf,Mathur:2011uj} and AMPS (firewall)  \cite{Almheiri:2012rt,Almheiri:2013hfa} constructions, which assume a single spacetime background. Consequently, speculative ideas such as black hole complementarity are not necessary to resolve the paradox: the final radiation state {\it encodes} the black hole quantum state, but the radiation modes remain independent degrees of freedom - they do not need to be identified with modes behind the horizon.

Euclidean wormholes (modulo unresolved issues with the path integral itself) allow one to directly calculate quantities such as the purity $\Tr[\, \rho_{ij}^2 \, ]$ of the radiation state. However, they do not allow one to calculate the quantum state of the radiation, and they do not directly reveal the physical, real time mechanism that allows the internal state of the black hole to influence Hawking radiation amplitudes. 

We discussed quantum hair on black holes, whose existence can be deduced using the Vilkovisky-DeWitt effective field theory in Minkowski spacetime \cite{Calmet:2021stu,Calmet:2021cip,Calmet:2019eof,Calmet:2022swf}. Quantum hair affects Hawking amplitudes \cite{Calmet:2023gbw}, and once its presence is accounted for we can explicitly follow the unitary time evolution of the quantum state of black hole plus radiation. Because this evolution is unitary, the entanglement entropy of the radiation follows the Page curve. At intermediate times the density matrix has the same non-sparse character as suggested by wormhole effects.

Both the quantum hair and replica wormhole pictures suggest that purity of the final radiation state is only recovered when one considers entanglement across all branches of a macroscopic superposition state.


\acknowledgments
The work of X.C. is supported in part  by the Science and Technology Facilities Council (grants numbers ST/T00102X/1 and ST/T006048/1).

\end{document}